\documentclass[a4paper,11pt]{article}
\usepackage{pos}
\usepackage{graphicx}
\usepackage{cancel}
\usepackage{empheq}
\usepackage{color}
\usepackage{hyperref}

\numberwithin{equation}{section}
\usepackage{float}
\restylefloat{table}
\restylefloat{figure}
\usepackage[utf8]{inputenc}
\usepackage[font=small, labelfont=bf]{caption}
\usepackage{multirow}
\usepackage{graphicx}
\usepackage{float}
\usepackage[numbers,sort&compress]{natbib}
\usepackage{enumerate}

\usepackage{graphicx}
\usepackage{cancel}
\usepackage{empheq}
\usepackage{color}

\usepackage{colortbl}
\definecolor{green2}{cmyk}{0, 1, 0.5, 0.3}
\definecolor{green3}{cmyk}{1, 0.75, 1.0, 0.0}

\definecolor{lightgreen}{cmyk}{0.2, 0, 0.2, 0.2}
\definecolor{lightgray}{cmyk}{0.1,0.2,0,0.1}
\definecolor{lightgray2}{cmyk}{0.4,0.4,0,0.8}
\definecolor{black}{cmyk}{1.0,1.0,1.0,1.0} 

\usepackage{hyperref}
\usepackage{float}
\restylefloat{table}
\restylefloat{figure}
\usepackage{longtable}
\usepackage{lipsum} 

\usepackage{comment}
\usepackage{graphicx}
\usepackage{dsfont}
\usepackage{float}
\usepackage{slashed}
\usepackage{setspace}
\usepackage[labelformat=simple]{subcaption}

\usepackage{pgfplots}
\usepackage{tikz}
\usepackage{tikz-cd}
\usetikzlibrary{shapes.misc,shadows,fit,decorations.pathmorphing,positioning,trees,decorations.markings,decorations.pathreplacing,calc,shapes,patterns,arrows,positioning,arrows.meta}
\usepackage{xcolor}
\usepackage{pgfplots}
\usepackage{pgfplotstable}
\usepackage{xcolor}
\usepackage{makecell}
\usepackage{diagbox}
\usepackage{environ}
\usepackage[utf8]{inputenc}

\usepackage{multirow}
\usepackage{graphicx}
\usepackage{float}
\usepackage[numbers,sort&compress]{natbib}
\usepackage{enumerate}
\usepackage{enumitem}
\usepackage[capitalize,sort]{cleveref}
\usepackage{hhline}
\usepackage{mathtools}
\usepackage{longtable}
\usepackage{makecell}
\usepackage{tabularx}
\usepackage{cellspace}
\usepackage{alphalph}
\usepackage{lscape}
\usepackage{slashed}
\usepackage{setspace}
\usepackage{pifont}

\crefname{figure}{Figure}{Figures}
\crefname{table}{Table}{Tables}

\def\be{\begin{equation}}
\def\ee{\end{equation}}
\def\bea{\begin{eqnarray}}
\def\eea{\end{eqnarray}}
\def\bes{\begin{subequations}}
	\def\ees{\end{subequations}}

\newcommand{\bmat}{\left(\begin{array}}
	\newcommand{\emat}{\end{array}\right)}

\def\ov{\overline}

\def\ov{\overline}
\def\1{{\bf 1}}
\def\2{{\bf 2}}
\def\3{{\bf 3}}
\def\4{{\bf 4}}
\def\6{{\bf 6}}

     \def\l{\lambda}

\newcommand{\nn}{\nonumber}

\newcommand{\beq}{\begin{equation}}
\newcommand{\eeq}{\end{equation}}

\def\ov{\overline}

\allowdisplaybreaks[2]
\numberwithin{equation}{section}

\def\l{\Lambda}

\def\be{\begin{equation}}
\def\ee{\end{equation}}
\def\bea{\begin{eqnarray}}
\def\eea{\end{eqnarray}}
\def\bes{\begin{subequations}}
	\def\ees{\end{subequations}}

\usepackage{multicol}
\usepackage{float}
\usepackage{caption}
\def\l   {{\langle}}

\allowdisplaybreaks[2]
\numberwithin{equation}{section}

\title{Perturbative LVS and Inflation: A Review of Volume Modulus and Fibre Scenarios}
\ShortTitle{Inflation in Perturbative LVS}

\author*[a]{George K. Leontaris}
\author[b]{Pramod Shukla}

\affiliation[a]{Physics Department, University of Ioannina\\
	University Campus, Ioannina 45110, Greece}

\affiliation[b]{Department of Physical Sciences, Bose Institute,\\
	Unified Academic Campus, EN 80, Sector V, Bidhannagar, Kolkata 700091, India}

\emailAdd{leonta@uoi.gr}
\emailAdd{pshukla@jcbose.ac.in}

\abstract{
In type IIB superstring compactifications, incorporating log-loop corrections and higher-derivative ${(\alpha^\prime)}^3$-corrections  can stabilize the overall volume of the compact internal space at exponentially large values. This mechanism forms the basis of the perturbative Large Volume Scenario (LVS). In this report, we briefly review two inflationary models realized within the perturbative LVS framework. The first one is the  volume modulus inflation (also known as inflection point inflation) and the second one is popularly known as fibre inflation. Using an explict Calabi-Yau orientifold, some concrete global embeddings of both the models are also discussed.}

\FullConference{Proceedings of the Corfu Summer Institute 2024 "School and Workshops on Elementary Particle Physics and Gravity" (CORFU2024)\
12 - 26 May, and 25 August - 27 September, 2024\\
Corfu, Greece\\}

\tableofcontents

\begin{document}
\maketitle


\section{Introduction}
\label{sec_intro}

Constructing realistic models of cosmological inflation within four-dimensional effective supergravities arising from string compactifications remains a major theoretical challenge. Over the past two decades, significant efforts have been devoted to this pursuit, leading to the development of various inflationary scenarios in this framework,
e.g. see \cite{Cicoli:2023opf,McAllister:2023vgy} and references therein.

A key ingredient in these model-building efforts is the existence of massless scalars referred to as moduli—ubiquitous in superstring compactifications. Remarkably, some of these moduli can serve as natural inflaton candidates when equipped with a sufficiently flat potential. However, before constructing viable string models, an essential first step (prerequisite to string model building) is moduli stabilization, since these fields determine the masses and couplings of the four-dimensional effective theory.

To realize single-field inflation, the moduli stabilization mechanism must be carefully designed to produce a mass hierarchy, ensuring one light modulus can naturally serve as the inflaton.
Type IIB superstring compactifications on Calabi-Yau orientifolds have emerged as particularly promising frameworks for this purpose, offering both theoretical appeal and practical viability,
given that several sources of scalar potential contributions are known, 
in terms of their dependencies on the various moduli. For example, the S-dual pair of the RR and NS-NS three-form fluxes, namely $F_3$ and $H_3$ can induce a perturbative superpotential \cite{Gukov:1999ya, Taylor:1999ii,Blumenhagen:2003vr} coupled with the axio-dilaton modulus $(S)$ and the complex structure moduli ($U^i$). However, the K\"ahler moduli $(T_\alpha)$ remain flat even in the presence of such background fluxes due to the so-called `no-scale structure'. Nevertheless, it turns out that the K\"ahler  moduli can also be stabilized once sub-leading corrections, such as the non-perturbative superpotential contributions \cite{Witten:1996bn,Green:1997di,Blumenhagen:2009qh}, and the ${\alpha'}$-corrections to the K\"ahler potential \cite{Becker:2002nn}, are included.

Following this route and the motivation, conventionally there have been two successful schemes for moduli stabilization proposed within the framework of type IIB superstring compactifications. These are the  KKLT scheme \cite{Kachru:2002sk} and the LARGE Volume Scenario (LVS) \cite{Balasubramanian:2005zx,Conlon:2005ki}. It turns out that the LVS scheme of moduli stabilization can dynamically stabilize the overall volume ${\cal V}$ of the compactifying CY threefold to exponentially large values. This is achieved by considering a combination of the perturbative ${\alpha^\prime}^3$ (BBHL) corrections to the K\"ahler potential ($K$) \cite{Becker:2002nn}, and the non-perturbative correction in the superpotential ($W$) \cite{Witten:1996bn,Green:1997di,Blumenhagen:2009qh}. Apart from the non-perturbative superpotential, the standard LVS needs the underlying CY threefold to possess a rigid diagonal del-Pezzo divisor to ensure the so-called `Swiss-cheese' structure in the volume-form. Further, the minimal LVS model is a two-field model realized with a CY threefold having $h^{1,1}_+({\rm CY}) = 2$ which fixes the overall volume ${\cal V}$ and the volume of the rigid four-cycle at leading order, and therefore models with $h^{1,1}_+({\rm CY}) \geq 3$ have been used to realize inflationary aspects in LVS. The main idea is that the third modulus, which remain flat at the level of the minimal LVS moduli stabilisation, can serve as inflaton candidate slowly rolling down a nearly flat potential. This framework has produced three distinct classes of inflationary models within the LVS paradigm, namely Blow-up inflation \cite{Conlon:2005ki,Blanco-Pillado:2009dmu,Cicoli:2017shd}, Fibre inflation \cite{Cicoli:2008gp,Cicoli:2016chb,Cicoli:2016xae,Cicoli:2017axo, Cicoli:2024bxw} and poly-instanton inflation \cite{Cicoli:2011ct,Blumenhagen:2012ue,Gao:2013hn,Gao:2014fva}.
Most recently, a novel variant called ``loop blow-up inflation" \cite{Bansal:2024uzr}, has been proposed, where the blow-up modulus acts as the inflaton field, with its potential generated entirely through string-loop corrections.

Both of these schemes, namely KKLT and LVS, utilize the non-perturbative superpotential contributions \cite{Witten:1996bn,Green:1997di} in order to stabilize the K\"ahler moduli, however such corrections may not be generically guaranteed in a given concrete setup. The reason being the fact that they depend on the specifics of the underlying CY geometry and the brane-setups/fluxes; these requirements can be like (i)~``Witten's unit arithmetic genus condition" \cite{Witten:1996bn} fulfilled for the CY having a rigid divisor, (ii)~``rigidification" of non-rigid divisors using magnetic fluxes on the divisors \cite{Bianchi:2011qh,Bianchi:2012pn,Louis:2012nb}, (iii)~visible sector and chirality issues \cite{Blumenhagen:2007sm,Blumenhagen:2008zz,Blumenhagen:2009qh,Cvetic:2012ts, Blumenhagen:2012kz}. For that reason, some alternate schemes of moduli stabilization have been proposed in the meantime which rely on using only the perturbative effects such as string-loop corrections \cite{Berg:2004ek,vonGersdorff:2005bf, Berg:2005ja, Berg:2005yu, Cicoli:2007xp, Gao:2022uop} or higher derivative F$^4$-corrections \cite{Ciupke:2015msa}, or the  non-geometric fluxes \cite{Aldazabal:2006up,deCarlos:2009fq,deCarlos:2009qm, Blumenhagen:2015kja,Shukla:2016xdy,Plauschinn:2020ram,Damian:2023ote}. 

In this context, another proposal has been made in \cite{Antoniadis:2018hqy,Antoniadis:2019rkh,Antoniadis:2020ryh} where it was shown that using logarithmic string loop corrections  (``log-loop" for short) to the K\"ahler potential along with the BBHL correction, one can realize an AdS minimum with exponentially large VEV for the overall volume ${\cal V}$ of the compactifying toroidal sixfold background. In fact, this proposal which has been initiated in toroidal setup has been further extended to a concrete K3-fibred CY setup in \cite{Leontaris:2022rzj} where it was shown that one can have $\langle {\cal V} \rangle \propto e^{c_1/g_s^2}$ where $g_s$ is the string coupling and $c_1 \simeq {\cal O}(1)$ positive constant. This is quite similar to the standard LVS but does not involve any non-perturbative effects, and hence this scheme is known as perturbative LVS. 

In the context of exploring the possibility of having some successful cosmological inflationary embedding in framework of perturbative LVS, some initiatives have been taken using the toroidal orientifold setups \cite{Antoniadis:2018ngr,Antoniadis:2019doc,Antoniadis:2020stf,Antoniadis:2021lhi}. Moreover, using the K3-fibred CY orientifold setup of \cite{Leontaris:2022rzj}, a couple of interesting single-field inflationary models have been recently proposed in \cite{Bera:2024ihl, Bera:2024zsk}. In this report, we plan to briefly review the following two models:

\noindent
{\bf Volume-modulus Inflation: } In the presence of suitable D-term effects, one can uplift the AdS minimum of the perturbative LVS to achieve de-Sitter minimum. Moreover, the overall volume modulus ${\cal V}$ is stabilized by a combination of two subleading effects, namely BBHL and log-loop correction to the K\"ahler potential, such that it is the lightest modulus, which can subsequently drive a small-field inflation, starting from a point close to the inflection point \cite{Antoniadis:2019doc,Antoniadis:2020stf}. For this reason, it is also referred to as ``inflection point inflation". A concrete global embedding of this proposal has been presented using a K3-fibred CY threefold in \cite{Bera:2024zsk} where the robustness of the models against various other corrections have been studied.

\noindent
{\bf Fibre Inflation: } This is a very popular large field inflationary model realized in the standard LVS by using a K3-fibred swiss-cheese CY threefold and a set of ``appropriate" string-loop corrections. The minimal construction of the fibre inflation consists of having three K\"ahler moduli corresponding to the CY volume of the form ${\cal V} = \lambda_f \tau_b \sqrt{\tau_f} - \lambda_s \tau_s^{3/2}$, where $\lambda$'s are some constants depending on the triple intersection numbers of the CY threefold with divisor volumes denoted as $\tau_b, \tau_f$ and $\tau_s$. Models based on K3-fibred CY$_3$ with $h^{1,1}({\rm CY}) \geq 3$ and having a diagonal dP divisor have been studied/classified in \cite{Cicoli:2016xae,Cicoli:2018tcq,Altman:2021pyc,Shukla:2022dhz,Crino:2022zjk}. The standard LVS fixes the overall volume ${\cal V}$ and the $\tau_s$ modulus which corresponds to the volume of the exceptional four-cycle needed to ensure the non-perturbative effects. The $\tau_f$ modulus corresponding to the volume of the K3 divisor remain flat at the leading order LVS, and receives some effective subleading corrections to the scalar potential via the  KK-type and winding-type string loop corrections \cite{Berg:2004sj,vonGersdorff:2005bf,Berg:2005ja,Berg:2007wt,Cicoli:2007xp,Gao:2022uop}. This subsequently helps in driving inflation by the $\tau_f$ modulus. However, it has been recently observed that despite being flat at the leading order LVS, the $\tau_f$ modulus is not free to move arbitrarily large distance in the moduli space due to severe restrictions arising from the K\"ahler cone conditions \cite{Cicoli:2017axo, Cicoli:2018tcq}.
These restrictions on the inflaton field range~\cite{Cicoli:2018tcq,Bera:2024ihl} originate from the exceptional divisor – an essential component of the standard LVS framework. This divisor plays a dual role: maintaining the Swiss-cheese structure in the Calabi-Yau volume form while simultaneously enabling non-perturbative superpotential contributions.
%
%
Given that perturbative LVS neither needs swiss-cheese structure or the non-perturbative superpotential contributions, it has been argued to alleviate this ``field-range" issue by embedding the fibre inflation machinery in the perturbative LVS framework \cite{Bera:2024ihl}.

\noindent
The review is organized as follows: In Section \ref{sec_preliminaries}, we present the necessary ingredients of the type IIB moduli stabilization regarding the standard LVS and the perturbative LVS.
In section \ref{sec_global} we present a concrete K3-fibre CY orientifold which resembles the toroidal case, and discuss the possible corrections to the effective scalar potential. Section \ref{sec_inflation} we review the two inflationary models with specific details on their single-field realization and robustness against additional corrections as present in the concrete global model. Finally, in section \ref{sec_summary} we summarize and conclude by giving some future directions.


\section{Relevant Preliminaries}
\label{sec_preliminaries}
In the context of moduli stabilization in type IIB superstring compactifications using the orientifolds of Calabi-Yau threefolds, the F-term scalar potential $V$ for the ${\cal N}=1$ effective four-dimensional theory is given by,
\be
\label{eq:V_gen}
e^{- {K}} \, V = {K}^{{\cal A} \ov {\cal B}} \, (D_{\cal A} W) \, (D_{\ov {\cal B}} \ov{W}) -3 |W|^2 \equiv V_{\rm cs} + V_{\rm k}\,,
\ee
where:
\be
\label{eq:VcsVk}
V_{\rm cs} =  K_{\rm cs}^{i \ov {j}} \, (D_i W) \, (D_{\ov {j}} \ov{W}) \qquad \text{and}\qquad V_{\rm k} =  K^{{A} \ov {B}} \, (D_{A} W) \, (D_{\ov {B}} \ov{W}) -3 |W|^2\,.
\ee
Here, $W$ denotes a holomorphic superpotential while $K$ denotes the K\"ahler potential which is a real function of the complexified moduli denoted as $\{{\cal A}, {\cal B}\} \in \{S, T_\alpha, U^i\}$. The covariant derivatives $D_{\cal A}$ are defined as $D_{\cal A}W = \partial_{\cal A}W + (K_{\cal A})W$. The moduli $\{S, T_\alpha, U^i\}$ are the ${\cal N} = 1$ ``chiral coordinates" obtained by complexifying various moduli with a set of RR axions. These are defined as: $S = c_0 + i\, s, \, \, T_\alpha = c_\alpha - i\, \tau_\alpha, \, \, U^i = v^i - i\, u^i$. Here, $s$ is the dilaton-dependent modulus, $u^i$'s are the complex structure saxions, and $\tau_\alpha$'s are the Einstein frame four-cycle volume moduli. Further, the $c_0$ and $c_\alpha$'s are universal RR axion and RR four-form axions, respectively, while the complex structure axions are denoted by $v^i$. Let us also note that the indices $\{i, \alpha\}$ are such that $i \in h^{2,1}_-({\rm CY}/{\cal O})$ while $\alpha \in h^{1,1}_+({\rm CY}/{\cal O})$. Also for the current review we assume that $h^{1,1} = h^{1,1}_+$ and hence, the  odd-moduli $G^a$ (e.g. see \cite{Cicoli:2021tzt}) are absent in our discussions.

The two functions $K$ and $W$ capture the low-energy dynamics of the four-dimensional effective supergravity theory. Depending on the possible corrections arising from the various sources, the K\"ahler potential and the superpotential induce useful contributions to the effective scalar potential. The schematic form of $K$ and $W$ can be given as
\be
\label{eq:K-W-gen}
K = -\ln\left[-i\int \Omega\wedge\bar{\Omega}\right]-\ln\left[-\,i\,(S-\bar{S})\right]-2\ln{\cal Y}, \qquad W= W_{\rm flux} + W_{\rm np}\,,
\ee
where $\Omega$ denotes the nowhere vanishing holomorphic 3-form of the compactifying Calabi-Yau threefold which depends on the complex-structure moduli, while ${\cal Y}$ encodes several contributions which mainly includes the CY volume ${\cal V}$ along with a series of other possible contributions such as a shift through the $\alpha'^3$ corrections \cite{Becker:2002nn}, also known as BBHL corrections, encoded in the parameter $\xi=-\frac{\chi(X)\,\zeta(3)}{2\,(2\pi)^3}$, where $\chi(X)$ is the CY Euler characteristic and $\zeta(3)\simeq 1.202$. For the superpotential, $W_{\rm flux}$ is induced by usual S-dual pair of the 3-form fluxes $(F_3, H_3)$ \cite{Gukov:1999ya} which depends on the $\{S, U^i\}$ moduli while the non-perturbative corrections $W_{\rm np}$ \cite{Witten:1996bn} can have $T_\alpha$ dependence to break the  no-scale symmetry needed to facilitate the volume moduli stabilization.  

In the context of 4D type IIB effective supergravity models, the conventional moduli stabilization is a two-step process. First, the complex structure moduli $U^i$ and the axio-dilaton $S$ are fixed by solving the following supersymmetric flatness conditions:
\bea
&& D_i W_{\rm flux} = 0 = D_{\ov {i}} \ov{W}_{\rm flux}, \qquad D_{S} W_{\rm flux} = 0 = D_{\ov {S}} \ov{W}_{\rm flux}.
\label{UStab}
\eea
The supersymmetric stabilization of $S$ and $U^i$ moduli leads to $\langle W_{\rm flux} \rangle = W_0$. The no-scale symmetry protects the K\"ahler moduli $T_\alpha$ which remain flat at the leading order. They can be stabilized in a second step via including other sub-leading contributions to the K\"ahler potential and/or the superpotential.


\subsection{Standard LVS}
The standard LVS scheme of moduli stabilization considers a combination of perturbative $(\alpha^\prime)^3$ corrections to the K\"ahler potential ($K$) and a non-perturbative contribution to the superpotential $(W)$ which can be generated by using rigid divisors, such as shrinkable del-Pezzo 4-cycles, or by rigidifying non-rigid divisors using magnetic fluxes \cite{Bianchi:2011qh, Bianchi:2012pn, Louis:2012nb}. The minimal LVS construction includes two K\"ahler moduli appearing in the Swiss-cheese like volume-form ${\cal V} =  \gamma_b \, \, \tau_{b}^{3/2} - \gamma_s\, \, \tau_{s}^{3/2}$ of the CY threefold, where $\gamma_b$ and $\gamma_s$ are determined through the triple intersection numbers on the CY threefold, and the 4-cycle volume moduli $\tau_\alpha$ are given by $\tau_\alpha = \partial_{t^\alpha} {\cal V}$ where  $t^{\alpha}$ denote the 2-cycle volume moduli.  However, for some unconventional scenarios, LVS moduli fixing can also be realized for special cases where the CY does not need to have a Swiss-cheese structure \cite{AbdusSalam:2020ywo}.

In order to realize the minimal LVS with two K\"ahler moduli, one needs the following ingredients
\bea
& & {\cal Y} = {\cal V}\,+\frac{\xi}{2}\left(\frac{S-\bar{S}}{2i}\right)^{3/2}, \quad W= W_0 + A_s\, e^{- i\, a_s\, T_s}\,,
\eea
where the presence of a `diagonal' del-Pezzo divisor, also referred to as `small' $4$-cycle of the CY threefold, induces the non-perturbative effect in the superpotential. After fixing $S$ and the $U$-moduli by imposing the supersymmetric flatness conditions, the flux superpotential $W_0$ and the pre-factor $A_s$ can effectively be considered as constant parameters. In fact, without any loss of generality, one can consider $W_0$ and $A_s$ to be a real quantities. Subsequently the leading order pieces in the large volume expansion are collected in three types of terms \cite{Balasubramanian:2005zx}:
\be
V \simeq \frac{\beta_{\alpha'}}{{\cal V}^3} + \beta_{\rm np1}\,\frac{\tau_s}{{\cal V}^2}\, e^{- a_s \tau_s} \cos\left(a_s \,c_s\right) \\
+ \beta_{\rm np2}\,\frac{ \sqrt{\tau_s}}{{\cal V}}\, e^{-2 a_s \tau_s},
\label{VlvsSimpl}
\ee
with:
\be
\beta_{\alpha'} = \frac{3 \,\kappa\,\hat\xi |W_0|^2}{4}\,, \quad \beta_{\rm np1} = 4 \, \kappa\, a_s |W_0| |A_s|\,, \quad \beta_{\rm np2} = 4 \, \kappa\, a_s^2 |A_s|^2 \sqrt{2 k_{sss}}, \quad \kappa = \frac{g_s\, e^{K_{cs}}}{8\pi}\,.
\ee
The conventional LVS scheme fixes the CY volume ${\cal V}$ along with the small divisor volume modulus $\tau_s$, and any LVS model with 3 or more K\"ahler moduli, $h^{1,1}({\rm CY})\geq 3$, can generically have flat directions at leading order. These flat directions can be promising inflaton candidates, depending on the geometric nature of the inflaton field and the source of inflaton potential.


\subsection{Perturbative LVS}

With the inclusion of the  log-loop corrections along with the BBHL corrections to the K\"ahler potential, one arrives at an effectively modified overall volume ${\cal V}$ which we denote as ${\cal Y} = {\cal Y}_0 + {\cal Y} _1$,
where ${\cal Y}_0$ denotes the overall volume modified by $\alpha^\prime$ corrections appearing at string tree-level while ${\cal Y}_1$ is induced at string 1-loop level as given  below  \cite{Antoniadis:2018hqy,Antoniadis:2018ngr,Antoniadis:2019doc,Antoniadis:2019rkh,Antoniadis:2020ryh,Antoniadis:2020stf},
\bea
& & {\cal Y}_0 = {\cal V} +  \frac{\xi}{2} \, e^{-\frac{3}{2} \phi} = {\cal V} + \frac{\xi}{2}\, \left(\frac{S-\ov{S}}{2\,{\rm i}}\right)^{3/2} \,, \\
& & {\cal Y}_1 = e^{\frac{1}{2} \phi}\, f({\cal V}) = \left(\frac{S-\ov{S}}{2\,{\rm i}}\right)^{-1/2} \left(\sigma + \eta \, \ln{\cal V}\right)\,.\nonumber
\label{eq:defY}
\eea
Here one has the following correlations among the various coefficients, $\xi$, $\sigma$ and $\eta$,
\bea
\label{eq:def-xi-eta}
& & \hskip-1cm  \xi = - \frac{\chi({\rm CY})\, \zeta[3]}{2(2\pi)^3}~, \qquad \sigma  = - \frac{\chi({\rm CY})\, \zeta[2]}{2(2\pi)^3} = - \, \eta, \qquad \frac{\xi}{\eta} = -\frac{\zeta[3]}{\zeta[2]} \\
& & \hskip-1cm \hat\xi = \frac{\xi}{g_s^{3/2}}~, \qquad \hat\eta = g_s^{1/2}\, \eta~, \qquad \qquad \frac{\hat\xi}{\hat\eta} = -\frac{\zeta[3]}{\zeta[2]\,g_s^2}~. \nonumber
\eea
This subsequently leads to the following form of the scalar potential,
\bea
\label{eq:masterV}
&&\hskip-1cm  V_{\rm pLVS} \simeq \frac{3\, \kappa\, \hat\xi}{4\, {\cal V}^3}\, |W_0|^2 + \frac{3 \, \kappa\, \hat\eta\, (\ln{\cal V} - 2)}{2{\cal V}^3}\,|W_0|^2 ~,
\eea
where as earlier, we have set $\kappa = \left(\frac{g_s\, e^{K_{cs}}}{8 \pi}\right)$. This scalar potential results in an exponentially large VEV for the overall volume determined by the following approximate relation \cite{Leontaris:2022rzj}:
\bea
\label{eq:pert-LVS}
& & \langle {\cal V} \rangle \simeq e^{-\frac{\hat\xi}{2\, \hat\eta} + \frac{7}{3}} = e^{a/g_s^2 + b}, \qquad a = \frac{\zeta[3]}{2 \zeta[2]} \simeq 0.365381, \quad b = \frac73~\cdot
\eea
Further, the Hessian analysis shows that one gets an AdS minimum with an exponentially large VEV of the overall volume $\cal V$. For numerical estimate one notes that $g_s = 0.2$ in Eq.~(\ref{eq:pert-LVS}) results in $\langle {\cal V} \rangle = 95593.3$. Subsequently, using the various possible uplifting methods, one can uplift this AdS minimum into a de-Sitter minimum, for example using D-term uplifting \cite{Burgess:2003ic}, anti-D3 uplifting \cite{Kachru:2002sk,Crino:2020qwk, Bento:2021nbb, AbdusSalam:2022krp,Cicoli:2024bxw} or T-brane uplifting \cite{Cicoli:2015ylx,Cicoli:2017shd,Cicoli:2021dhg}.


\section{A toroidal-like Calabi-Yau orientifold}
\label{sec_global}
Motivated by the proposal of \cite{Antoniadis:2018hqy,Antoniadis:2018ngr,Antoniadis:2019doc,Antoniadis:2019rkh,Antoniadis:2020ryh,Antoniadis:2020stf, Antoniadis:2021lhi} where some underlying symmetries of the compactifying toroidal orientifold have been found to be useful for realizing the perturbative LVS, we begin by presenting an explicit Calabi-Yau threefold which possesses a toroidal-like volume form, i.e. ${\cal V} \propto \sqrt{\tau_1, \tau_2 \tau_3}$. For this purpose, the CY dataset of Kreuzer-Skarke \cite{Kreuzer:2000xy} with $h^{1,1} = 3$ was scanned and it was found that there are a couple of geometries which could suitably give this volume form \cite{Gao:2013pra,Leontaris:2022rzj}.  

\subsection{Toric data}
We consider a CY threefold corresponding to the polytope Id: 249 in the CY database of \cite{Altman:2014bfa} can be defined by the following toric data:
\begin{center}
\begin{tabular}{|c|ccccccc|}
\hline
\cellcolor[gray]{0.9}Hyp &\cellcolor[gray]{0.9} $x_1$  &\cellcolor[gray]{0.9} $x_2$  &\cellcolor[gray]{0.9} $x_3$  &\cellcolor[gray]{0.9} $x_4$  &\cellcolor[gray]{0.9} $x_5$ & \cellcolor[gray]{0.9}$x_6$  &\cellcolor[gray]{0.9} $x_7$       \\
\hline
\cellcolor[gray]{0.9}4 & 0  & 0 & 1 & 1 & 0 & 0  & 2   \\
\cellcolor[gray]{0.9}4 & 0  & 1 & 0 & 0 & 1 & 0  & 2   \\
\cellcolor[gray]{0.9}4 & 1  & 0 & 0 & 0 & 0 & 1  & 2   \\
\hline
& $K3$  & $K3$ & $K3$ &  $K3$ & $K3$ & $K3$  &  SD  \\
\hline
\end{tabular}
\end{center}
The Hodge numbers are $(h^{2,1}, h^{1,1}) = (115, 3)$, the Euler number is $\chi=-224$ and the Stanley-Reisner ideal is:
\be
{\rm SR} =  \{x_1 x_6, \, x_2 x_5, \, x_3 x_4 x_7 \} \,. \nn
\ee
We also note that this CY threefold was encountered earlier, e.g. while exploring the possibility of including the odd-moduli via the exchange of the  non-trivially identical divisors in \cite{Gao:2013pra}. Moreover, an upgraded version of this CY has been used for the chiral global embedding of Fibre inflation model \cite{Cicoli:2017axo}.

Using {\it cohomCalg} \cite{Blumenhagen:2010pv, Blumenhagen:2011xn}, the various divisor topologies turn out to be encoded in the following Hodge diamonds:
\bea
K3 &\equiv& \begin{tabular}{ccccc}
    & & 1 & & \\
   & 0 & & 0 & \\
  1 & & 20 & & 1 \\
   & 0 & & 0 & \\
    & & 1 & & \\
  \end{tabular}, \qquad \quad {\rm SD} \equiv \begin{tabular}{ccccc}
    & & 1 & & \\
   & 0 & & 0 & \\
  27 & & 184 & & 27 \\
   & 0 & & 0 & \\
    & & 1 & & \\
  \end{tabular}.
\eea
Moreover, the curves at the intersection loci of two generic coordinate divisors are given in Table \ref{Tab1} which shows that all the K3 divisors interest with one another on a ${\mathbb T}^2$. This is precisely what one has for the standard ${\mathbb T}^6 = {\mathbb T}^2 \otimes {\mathbb T}^2 \otimes {\mathbb T}^2$ case. These symmetries are consistent with the basic requirement for generating logarithmic string-loop effects as elaborated in \cite{Antoniadis:2018hqy,Antoniadis:2018ngr,Antoniadis:2019doc,Antoniadis:2019rkh,Antoniadis:2020ryh,Antoniadis:2020stf,Antoniadis:2021lhi}.
\begin{table}[h]
  \centering
 \begin{tabular}{|c|c|c|c|c|c|c|c|}
\hline
\cellcolor[gray]{0.9}  &\cellcolor[gray]{0.9} $D_1$  &\cellcolor[gray]{0.9} $D_2$  &\cellcolor[gray]{0.9} $D_3$  & \cellcolor[gray]{0.9}$D_4$  & \cellcolor[gray]{0.9}$D_5$ &\cellcolor[gray]{0.9} $D_6$  & \cellcolor[gray]{0.9}$D_7$  \\
    \hline
		\hline
\cellcolor[gray]{0.9}$D_1$ & $\emptyset$  &  ${\mathbb T}^2$      &  ${\mathbb T}^2$        &  ${\mathbb T}^2$   &  ${\mathbb T}^2$  &  $\emptyset$   &  ${\cal H}_9$  \\
\cellcolor[gray]{0.9}$D_2$ &  ${\mathbb T}^2$ & $\emptyset$        &  ${\mathbb T}^2$        &  ${\mathbb T}^2$   &    $\emptyset$   & ${\mathbb T}^2$  & ${\cal H}_9$
\\
\cellcolor[gray]{0.9}$D_3$  &  ${\mathbb T}^2$      &  ${\mathbb T}^2$        & $\emptyset$ & $\emptyset$ &  ${\mathbb T}^2$   &  ${\mathbb T}^2$   &  ${\cal H}_9$
\\
\cellcolor[gray]{0.9}$D_4$  &  ${\mathbb T}^2$      &  ${\mathbb T}^2$        & $\emptyset$ & $\emptyset$ &  ${\mathbb T}^2$   &  ${\mathbb T}^2$   &  ${\cal H}_9$
\\
\cellcolor[gray]{0.9}$D_5$ &  ${\mathbb T}^2$ & $\emptyset$        &  ${\mathbb T}^2$        &  ${\mathbb T}^2$   &    $\emptyset$   & ${\mathbb T}^2$  & ${\cal H}_9$
\\
\cellcolor[gray]{0.9}$D_6$ & $\emptyset$  &  ${\mathbb T}^2$      &  ${\mathbb T}^2$        &  ${\mathbb T}^2$   &  ${\mathbb T}^2$  &  $\emptyset$   &  ${\cal H}_9$  \\
\cellcolor[gray]{0.9}$D_7$ & ${\cal H}_9$  &  ${\cal H}_9$      &  ${\cal H}_9$        &  ${\cal H}_9$   &  ${\cal H}_9$  &  ${\cal H}_9$   &  ${\cal H}_{97}$
\\
    \hline
  \end{tabular}
  \caption{Intersection curves of the two coordinate divisors. Here ${\cal H}_g$ denotes a curve with Hodge numbers $h^{0,0} = 1$ and $h^{1,0} = g$, and hence ${\cal H}_1 \equiv {\mathbb T}^2$, while ${\cal H}_0 \equiv {\mathbb P}^1$.}
\label{Tab1}
\end{table}

\noindent
Considering the basis of smooth divisors $\{D_1, D_2, D_3\}$ we get the following intersection polynomial which has just one non-zero classical triple intersection number \footnote{There is another CY threefold in the database of \cite{Altman:2014bfa} which has the intersection polynomial of the form $I_3 = D_1 D_2 D_3$, however that CY threefold (corresponding to the polytope Id: 52) has non-trivial fundamental group.}:
\bea
& & I_3 = 2\, D_1\, D_2\, D_3,
\eea
while the second Chern-class of the CY is given by,
\bea
c_2({\rm CY}) = 5 D_3^2+12 D_1 D_2 + 12 D_2 D_3+12 D_1 D_3.
\eea
Subsequently, considering the K\"ahler form $J = \sum_{\alpha =1}^3 t^\alpha D_\alpha$, the overall volume and the 4-cycle volume moduli can be given as follows:
\bea
& & \hskip-1cm {\cal V} = 2\, t^1\, t^2\, t^3, \qquad \qquad \tau_1 = 2\, t^2 t^3,  \quad  \tau_2 = 2\, t^1 t^3, \quad  \tau_3 = 2 \,t^1 t^2 \,.
\label{Taus}
\eea
This volume form can also be expressed as follows:
\bea
& & \hskip-1cm {\cal V} = 2 \, t^1\, t^2\, t^3  = \frac{1}{\sqrt{2}}\,\sqrt{\tau_1 \, \tau_2\, \tau_3}~.
\eea
This demonstrates that the volume form ${\cal V}$ exhibits toroidal characteristics  with an exchange symmetry $1 \leftrightarrow 2 \leftrightarrow 3$ under which all the three $K3$ divisors which are part of the basis are exchanged. Further, the K\"ahler cone for this setup is described by the conditions below,
\bea
\label{KahCone}
\text{K\"ahler cone:} &&  t^1 > 0\,, \quad t^2 > 0\,, \quad t^3 > 0\,.
\eea
We note that the volume form can also be expressed as,
\bea
\label{eq:t-tau-vol}
{\cal V} = t^1 \, \tau_1 = t^2 \, \tau_2  = t^3 \, \tau_3,
\eea
which means that the transverse distance for the stacks of $D7$-branes wrapping the divisor $D_1$ is given by $t^1$ and similarly $t^2$ is the transverse distance for $D7$-branes wrapping the divisor $D_2$ and so on. Furthermore, we note that the second Chern numbers, namely $\Pi_\alpha$, corresponding to each of the seven coordinate divisors are given as,
\bea
& & \Pi_\alpha = 24 \quad \forall \, \alpha \in \{1, 2,..,6\}; \quad \Pi_7 = 124.
\eea
Finally, the tree-level metric for the K\"ahler moduli space takes the following form,
\bea
\label{eq:Kij-tree}
& & K_{\alpha\beta}^0 = \frac{1}{4\, {\cal V}^2} \left(
\begin{array}{ccc}
 (t^1)^2 & 0 & 0 \\
 0 & (t^2)^2 & 0 \\
 0 & 0 & (t^3)^2 \\
\end{array}
\right) = \frac{1}{4} \left(
\begin{array}{ccc}
 (\tau_1)^{-2} & 0 & 0 \\
 0 & (\tau_2)^{-2} & 0 \\
 0 & 0 & (\tau_3)^{-2} \\
\end{array}
\right),
\eea
where we have used (\ref{eq:t-tau-vol}) in the second step.


\subsection{Orientifold involution, fluxes and brane setting}
For a given holomorphic involution, one needs to introduce D3/D7-branes and fluxes in order to cancel all the charges. For example, one can nullify the D7-tadpoles via introducing stacks of $N_a$ D7-branes wrapped around suitable divisors (say $D_a$) and their orientifold images ($D_a^\prime$) such that the following relation holds \cite{Blumenhagen:2008zz}:
\bea
\label{eq:D7tadpole}
& & \sum_a\, N_a \left([D_a] + [D_a^\prime] \right) = 8\, [{\rm O7}]\,.
\eea
Moreover, the presence of D7-branes and O7-planes also contributes to the D3-tadpoles, which, in addition, receive contributions from  $H_3$ and $F_3$ fluxes, D3-branes and O3-planes. The D3-tadpole cancellation condition is given as \cite{Blumenhagen:2008zz}:
\be
N_{\rm D3} + \frac{N_{\rm flux}}{2} + N_{\rm gauge} = \frac{N_{\rm O3}}{4} + \frac{\chi({\rm O7})}{12} + \sum_a\, \frac{N_a \left(\chi(D_a) + \chi(D_a^\prime) \right) }{48}\,,
\label{eq:D3tadpole}
\ee
where $N_{\rm flux} = (2\pi)^{-4} \, (\alpha^\prime)^{-2}\int_X H_3 \wedge F_3$ is the contribution from background fluxes and $N_{\rm gauge} = -\sum_a (8 \pi)^{-2} \int_{D_a}\, {\rm tr}\, {\cal F}_a^2$ is due to D7 worldvolume fluxes. However, for the simple case where D7-tadpoles are cancelled by placing 4 D7-branes (plus their images) on top of an O7-plane, the condition (\ref{eq:D3tadpole}) reduces to the following form:
\be
N_{\rm D3} + \frac{N_{\rm flux}}{2} + N_{\rm gauge} =\frac{N_{\rm O3}}{4} + \frac{\chi({\rm O7})}{4}\,.
\label{eq:D3tadpole1}
\ee
It turns out that the involution $x_7 \to - x_7$ yields a favorable brane configuration. This results in only one fixed point set with $\{O7 = D_7\}$ along with no $O3$-planes, and subsequently one can consider the brane setting having three stacks of $D7$-branes wrapping each of the three divisors $\{D_1, D_2, D_3\}$ in the basis,
\bea
& & 8\, [O_7] = 8 \left([D_1] + [D_1^\prime] \right) + 8 \left([D_2] + [D_2^\prime] \right) + 8 \left([D_3] + [D_3^\prime] \right)\,,
\eea
along with the $D3$ tadpole cancellation condition being given as
\be
N_{\rm D3} + \frac{N_{\rm flux}}{2} + N_{\rm gauge} = 0 + \frac{240}{12} + 8 + 8 + 8 = 44\,.
\ee
Therefore, the current CY orientifold construction limits the net number of D3-brane charges to $N_3 = 44$ following from the D3 tadpole constraints and hence $|\hat{W}_0| \lesssim 6$ \cite{Denef:2004cf,Denef:2004ze}. However a different CY orietifold may relax this condition significantly, e.g. those with large $N_3$ \cite{Crino:2022zjk,Shukla:2022dhz}.


\subsection{Possible sub-leading corrections to the scalar potential}
Given that there are no rigid divisors present, a priori this setup will not receive non-perturbative superpotential contributions from instanton or gaugino condensation. However, there can be various perturbative contributions such as the BBHL correction, and KK/winding-type string-loop corrections which are given as below, \cite{Berg:2004sj,vonGersdorff:2005bf,Berg:2005ja,Berg:2007wt,Cicoli:2007xp,Gao:2022uop}
\bea
\label{eq:Vgs-KK-Winding}
& & V_{g_s}^{\rm KK} = \kappa \, g_s^2 \, \frac{|W_0|^2}{16\,{\cal V}^4} \sum_{\alpha,\beta} C_\alpha^{\rm KK} C_\beta^{\rm KK} \left(2\,t^\alpha t^\beta - 4\, {\cal V} \,k^{\alpha\beta}\right), \nonumber\\
& & V_{g_s}^{\rm W} = - \frac{\kappa\,|W_0|^2}{{\cal V}^3} \, \sum_{\alpha=1}^3 \frac{C_\alpha^W}{t^\alpha}\,,
\eea
where $C_\alpha^{\rm KK}$ and $C_\alpha^W$ are some model dependent coefficients which can generically depend on the CS moduli. The divisor intersection curves in Table \ref{Tab1} show that all the three $D7$-brane stacks intersect at ${\mathbb T}^2$ while each of those intersect the $O7$-plane on a curve ${\cal H}_9$ defined by $h^{0,0} = 1$ and $h^{1,0} = 9$. These properties about the transverse distances and the divisor interesting on ${\mathbb T}^2$ is perfectly like what one has for the toroidal case, though the divisors are $K3$ for the current situation unlike ${\mathbb T}^4$ divisors of the six-torus. These symmetries are consistent with the basic requirement for generating logarithmic string-loop effects as elaborated in \cite{Antoniadis:2018hqy,Antoniadis:2018ngr,Antoniadis:2019doc,Antoniadis:2019rkh,Antoniadis:2020ryh,Antoniadis:2020stf,Antoniadis:2021lhi}.

Further we note that there are no non-intersecting $D7$-brane stacks and the $O7$-planes along with no $O3$-planes present as well, and therefore this model does not induce the KK-type string-loop corrections to the K\"ahler potential. However, given the fact that each of the three $D7$-brane stacks as well as $O7$-plane intersect one another on non-contractible curves (e.g. see Table \ref{Tab1}), one will have string-loop effects of the winding-type to be given as below \cite{Bera:2024zsk},
\bea
\label{eq:Vgs-Winding-globalmodel}
& & V_{g_s}^{\rm W} = - \frac{\kappa\,|W_0|^2}{{\cal V}^3} \left(\frac{C_{w_1}}{t^1} + \frac{C_{w_2}}{t^2} +\frac{C_{w_3}}{t^3} +\frac{C_{w_4}}{2(t^1+t^2)} +\frac{C_{w_5}}{2(t^2+t^3)} +\frac{C_{w_6}}{2(t^3+t^1)} \right)\,,
\eea
where $C_{w_i}$'s are complex-structure moduli dependent quantities and can be taken as parameter for the moduli dynamics of the sub-leading effects. Further, let us also note that although this CY have several properties like a toroidal case, the divisor being $K3$ implies that $\Pi = 24$. This is unlike the six-torus case where the ${\mathbb T}^4$ divisor has a vanishing $\Pi$, and hence no higher derivative F$^4$ effects. For our case, we find the following corrections to the scalar potential ,
\bea
\label{eq:F^4-term-globalmodel}
& & V_{{\rm F}^4} \equiv - \frac{\lambda\,\kappa^2\,|W_0|^4}{g_s^{3/2} {\cal V}^4}\, \Pi_\alpha t^\alpha = - \frac{\lambda\,\kappa^2\,|W_0|^4}{g_s^{3/2} {\cal V}^4}\, 24 \, \left(t^1 + t^2 + t^3\right).
\eea
In addition, by appropriately turning on the worldvolume gauge fluxes ${\cal F}$ on the stacks of D7-brane wrapping the K3 divisors, one can generate the following D-term contributions to the scalar potential \cite{Bera:2024zsk},
\bea
\label{eq:VDup2}
& & V_{\rm D} = \frac{d_1}{\tau_1} \left(\frac{q_{12}}{\tau_2} + \frac{q_{13}}{\tau_3}\right)^2 + \frac{d_2}{\tau_2} \left(\frac{q_{21}}{\tau_1} + \frac{q_{23}}{\tau_3}\right)^2 + \frac{d_3}{\tau_3} \left(\frac{q_{31}}{\tau_1} + \frac{q_{32}}{\tau_2}\right)^2,
\eea
where 
\bea
& & q_{\alpha\beta} = \int_{\rm CY} D_\alpha \wedge D_\beta \wedge {\cal F}.
\eea
Such D-term contributions can be used for uplifting the AdS minimum into a de-Sitter minimum. However, in the absence of such gauge fluxes on the divisors wrapping the D7-stacks, other uplifting sources can also be used, e.g. T-brane uplifting, and anti-D3 uplifting. 


\section{Inflation in perturbative LVS}
\label{sec_inflation}
\noindent
In summary, the effective scalar potential takes the following form
\bea
\label{eq:Vfinal-simp}
& & \hskip-1cm V_{\rm tot} \simeq V_{\rm up} + \, {\cal C}_1 \, \left(\frac{\hat\xi - 4\,\hat\eta + 2\,\hat\eta \, \ln{\cal V}}{{\cal V}^3}\right) \\
& &  \hskip-0.5cm - \frac{{\cal C}_2}{{\cal V}^4} \left(C_{w_1}\tau_1 + C_{w_2}\tau_2 + C_{w_3}\tau_3 + \frac{C_{w_4}\tau_1 \tau_2}{2(\tau_1 + \tau_2)} + \frac{C_{w_5}\tau_2 \tau_3}{2(\tau_2 + \tau_3)} + \frac{C_{w_6}\tau_3 \tau_1}{2(\tau_3 + \tau_1)}\right) \nonumber\\
& & \hskip-0.5cm +  \frac{{\cal C}_3}{{\cal V}^3}\,\left(\frac{1}{\tau_1} + \frac{1}{\tau_2}+\frac{1}{\tau_3} \right), \nonumber
\eea
where the various coefficients ${\cal C}_i$'s are given by,
\bea
\label{eq:calCis}
& & \hskip-1cm {\cal C}_1 = \frac{3\,\kappa\, |W_0|^2}{4}, \quad {\cal C}_2 = \frac{4\, {\cal C}_1}{3}, \quad {\cal C}_3 = - \frac{24\, \lambda\,\kappa^2\, |W_0|^4}{g_s^{3/2}}, \quad |\lambda| =  \, {\cal O}(10^{-4}), \quad \kappa = \frac{g_s\, e^{K_{cs}}}{8\pi}. \nonumber
\eea

\begin{itemize}

\item
The first line of Eq.~(\ref{eq:Vfinal-simp}) has two contributions; the first one (i.e. $V_{\rm up}$) stands for  the uplifting pieces and the other one is a combination of BBHL and string log-loop effects. The first term can be schematically taken as: $V_{up} \propto {\cal V}^{-n}$ for $n = 4/3, 2, 8/3$ corresponding to anti-D3 uplifting, D-term uplifting  and T-brane uplifting. The second piece appears at ${\cal O}({\cal V}^{-3})$ in the large volume expansion.

\item
The second line of Eq.~(\ref{eq:Vfinal-simp}) presents the typical winding type string-loop effects which appears at ${\cal O}({\cal V}^{-10/3})$ in the large volume expansion. In fact there can be additional loop corrections motivated by the field theoretic computations \cite{vonGersdorff:2005bf,Gao:2022uop}, however we do not include those corrections in the current discussions.

\item
The third line of Eq.~(\ref{eq:Vfinal-simp}) presents the higher derivative F$^4$-corrections which appear at ${\cal O}({\cal V}^{-11/3})$ in the large volume expansion.

\end{itemize}
\noindent
Now we will show that 
\begin{enumerate}
\item 
one can realize the  volume modulus inflation, also known as ``inflection point inflation" using the first line of (\ref{eq:Vfinal-simp}). In this case, one has to ensure that the inflationary dynamics is stable against the contributions in the second and third line of (\ref{eq:Vfinal-simp}).

\item
one can realize the fibre inflation using the second and third line of (\ref{eq:Vfinal-simp}) via keeping the overall volume fixed using the perturbative LVS.

\end{enumerate} 

\noindent
The slow-roll parameters are generically defined through the derivatives of the Hubble parameter:
\bea
\label{eq:slow-roll-H}
& & \hskip-1.5cm \epsilon_H = -\frac{\dot{H}}{H^2} = \frac{1}{H} \frac{d H}{dN_e}, \qquad \qquad \eta_H = \frac{\dot\epsilon_H}{\epsilon_H \, H} = \frac{1}{\epsilon_H} \frac{d\epsilon_H}{dN_e},
\eea
where dot denotes the time derivative while $N_e$ denotes the number of e-foldings determined by,
\bea
\label{eq:e-fold-def}
& & N_e(\phi) = \int H \, dt = \int_{\phi_{\rm end}}^{\phi_\ast} \, \frac{1}{\sqrt{2 \epsilon_H}}\, d\phi \, \simeq \, \int_{\phi_{\rm end}}^{\phi_\ast} \, \frac{V_{\rm inf}}{V^\prime_{\rm inf}}\, d\phi\,\,,
\eea
where $\phi_\ast$ is  the point of horizon exit at which the cosmological observables are to be matched with the experimentally observed values. However, the slow-roll inflationary parameters can  also be defined through the derivatives of the potential
\bea
\label{eq:slow-roll-V-def}
& & \epsilon_V \equiv \frac{1}{2} \left(\frac{V^\prime_{\rm inf}}{V_{\rm inf}}\right)^2 , \qquad \eta_V \equiv \frac{V^{\prime\prime}_{\rm inf}}{V_{\rm inf}}. \nonumber
\eea
In fact, for single field inflation, the two sets of slow-roll parameters, namely $(\epsilon_H, \eta_H)$ and $(\epsilon_V, \eta_V)$ can be correlated as $\epsilon_V \simeq \epsilon_H, \, \eta_H \simeq -2\, \eta_V + 4\, \epsilon_V$ (e.g. see \cite{Achucarro:2018vey}), and subsequently the cosmological observables such as the scalar perturbation amplitude, the spectral index, and the tensor-to-scalar ratio are correlated with the slow-roll parameters $\epsilon_V$ and $\eta_V$ as below \cite{Planck:2018jri},
\bea
\label{eq:cosmo-observables}
& & P_s \equiv \frac{V_{\rm inf}^\ast}{24 \pi^2 \, \epsilon_H^\ast} \simeq 2.1 \times 10^{-9}, \qquad {\rm or} \qquad \frac{V_{\rm inf}^{\ast3}}{V_{\rm inf}^{\prime\ast^2}} \simeq 2.6\times 10^{-7},\\
& & n_s - 1 = -2 \epsilon_H^\ast - \eta_H^\ast \simeq 2 \, \eta_V^\ast - 6\, \epsilon_V^\ast \simeq -0.04,\nonumber\\
& & r = 16 \epsilon_H^\ast \simeq 16 \epsilon_V^\ast,\nonumber
\eea
where all the cosmological observables are evaluated at the horizon exit $\phi = \phi^\ast$ and one also has sufficient e-foldings: $N_e(\phi^\ast) \gtrsim 60$. In fact, the number of e-foldings $N_e$ depends on many things including the post-inflationary aspects and can be given as a sum several contributions \cite{Liddle:2003as,Cicoli:2017axo}:
\bea
\label{eq:cosmo-observables1}
& & \hskip-0.5cm N_e \simeq \int_{\phi_{\rm end}}^{\phi_\ast} \frac{V}{V^\prime} d\phi \simeq 57 + \frac{1}{4} \ln(r_\ast V_\ast) - \frac{1}{3}\ln\left(\frac{10V_{\rm end}}{m_{\inf}^{3/2}}\right),\nonumber
\eea
where $\phi_{\rm end}$ corresponds to end of inflation determined by $\epsilon_H = 1$ and $m_{\rm inf}$ is the inflaton mass. Also, it is worth noting that typically $N_e \simeq 50$ for Fibre inflation \cite{Cicoli:2017axo,Bhattacharya:2020gnk,Cicoli:2020bao}.

\subsection{Inflection-Point Inflation}
\label{sec_volInflation}
Considering the tree-level K\"ahler metric arising from the volume form ${\cal V} = \frac{\sqrt{\tau_1\tau_2\tau_3}}{\sqrt{n_0}}$, one obtains a set of canonically normalized fields $\varphi^\alpha$ related to the $4$-cycle volume moduli $\{\tau_1, \tau_2, \tau_3\}$ via the following relations,
\bea
\label{eq:cononical-varphi0}
& & \varphi^\alpha = \frac{1}{\sqrt{2}} \ln \tau_\alpha, \qquad \forall \, \alpha \in \{1, 2, 3\}.
\eea 
Given that the overall volume ${\cal V}$ serves as a good expansion parameter for a series of possible perturbative corrections, it is useful to define the following set of canonical normalized fields $\phi^\alpha$,
\bea
\label{eq:cononical-varphi1}
& & \phi^1 = \frac{1}{\sqrt{3}} \left(\varphi^1+ \varphi^2 + \varphi^3 \right) = \sqrt{\frac{2}{3}} \ln(\sqrt{n_0}\,{\cal V}) , \\
& & \phi^2 = \frac{1}{\sqrt{2}} \left(\varphi^1- \varphi^2 \right), \qquad \phi^3 = \frac{1}{\sqrt{6}} \left(\varphi^1+ \varphi^2  -2 \varphi^3 \right). \nonumber
\eea
This choice of cannonical fields is further motivated by the isotropic considerations. In particular, for isotropic moduli stabilization with $\langle \varphi^1 \rangle \simeq \langle \varphi^2 \rangle \simeq \langle \varphi^3 \rangle$,  one obtains
$\langle \phi^2 \rangle \simeq \langle \phi^3 \rangle \simeq 0$. Using (\ref{eq:cononical-varphi1}), the total scalar potential in Eq.~(\ref{eq:Vfinal-simp}) takes the following form,
\bea
\label{eq:Vgen2}
& & \hskip-1.5cm V = e^{-\sqrt{6}\phi^1}\, \biggl[d_1 e^{-\, \phi^2 - \sqrt{3} \, \phi^3} \left(q_{12} \,e^{\phi^2} + q_{13} \, e^{\sqrt{3} \phi^3} \right)^2 \\
& & + d_2 e^{-\, \phi^2 - \sqrt{3} \, \phi^3} \left(q_{21} + q_{23} \, e^{\phi^2 + \sqrt{3} \phi^3} \right)^2 + d_3 e^{-2\, \phi^2} \left(q_{31} \,+ q_{32} \, e^{2\phi^2} \right)^2 \biggr] \nonumber\\
& & \hskip-1cm + 2\, \hat\eta\, e^{-3\sqrt{\frac{3}{2}} \phi^1}\, n_0^{3/2} \, {\cal C}_1\, \left(\sqrt{\frac{3}{2}} \, \phi^1 +\frac{\hat\xi}{2\,\hat\eta} - 2 - \frac{1}{2} \, \ln n_0 \right) \nonumber\\
& & \hskip-1cm + n_0^2 \, {\cal C}_2 \, {\cal C}_w \,e^{-5\sqrt{\frac{2}{3}} \phi^1} \biggl(e^{-\frac{2}{\sqrt{3}} \phi^3}+ e^{-\phi^2 + \frac{1}{\sqrt{3}}\phi^3} + e^{\phi^2 + \frac{1}{\sqrt{3}}\phi^3} + \frac{e^{\phi^2 + \frac{1}{\sqrt{3}}\phi^3}}{2(1+\,e^{2\phi^2})} \nonumber\\
& & \hskip-0.0cm + \frac{e^{\frac{1}{\sqrt{3}}\phi^3}}{2(e^{\phi^2}+e^{\sqrt{3} \phi^3})} + \frac{e^{\phi^2 + \frac{1}{\sqrt{3}}\phi^3}}{2(1+\,e^{\phi^2+\sqrt3 \phi^3})}  \biggr), \nonumber\\
& & \hskip-1cm + n_0^{3/2} \, {\cal C}_3 \,e^{-\frac{11}{\sqrt{6}} \phi^1}  \left(e^{-\phi^2 - \frac{1}{\sqrt{3}}\phi^3} + e^{\phi^2 - \frac{1}{\sqrt{3}}\phi^3} + e^{\frac{2}{\sqrt{3}} \phi^3}\right). \nonumber
\eea
Here we have set the $C_{w_i}$ parameters as $C_{w_1} = C_{w_2} = C_{w_3} = C_{w_4} = C_{w_5} = C_{w_6} \equiv -\, {\cal C}_w$. Note that (\ref{eq:Vgen2}) shows that the $\phi^1$ dependence appears as an overall factor in $V_{\rm up}$ despite its form being rather complicated. Using the generic scalar potential in Eq.~(\ref{eq:Vgen2}), the three extremisation conditions arising from $\partial_{\phi^\alpha} V = 0$ can be equivalently expressed as,
\bea
\label{eq:V-task2-3}
& & \hskip-1cm d_1 =  - {\cal Q}\,\biggl[\hat\eta \, {\cal C}_1\, n_0^{3/2}\, e^{-\sqrt{\frac{3}{2}} \langle\phi^1\rangle} \left(\sqrt{\frac{3}{2}} \, \langle \phi^1 \rangle -a_2\right) +\frac{25}{12} \, n_0^2 \, {\cal C}_2 \, {\cal C}_w \,e^{-2\sqrt{\frac{2}{3}} \langle\phi^1\rangle} \\
& & \hskip-0.5cm  + \frac{11}{6} \, n_0^{3/2} \, {\cal C}_3 \, e^{-\frac{5}{\sqrt6} \langle\phi^1\rangle}\biggr], \qquad \langle \phi^2 \rangle = 0 = \langle \phi^3 \rangle, \nonumber
\eea
subject to simultaneously satisfying the following relations,
\bea
\label{eq:d-qs1}
& & \hskip-1cm d_2 = d_1 \frac{q_{12}^2-q_{13}^2}{q_{23}^2-q_{21}^2} > 0, \quad d_3 = d_1 \frac{q_{12}^2-q_{13}^2}{q_{31}^2-q_{32}^2} > 0, \quad a_2 = -\frac{\hat\xi}{2\,\hat\eta} + \frac{7}{3} + \frac{1}{2} \, \ln n_0 \, > \, 0.
\eea
Here, ${\cal Q} \neq 0$ is a ratio depending on the flux parameters $q_{\alpha\beta}$'s which can be given as,
\bea
\label{eq:Rq}
& & {\cal Q}^{-1} = \frac{(q_{12}+q_{13})}{3(q_{21}-q_{23})(q_{31}-q_{32})}\biggl(q_{13}q_{21}(q_{31}-3q_{32})+q_{12}q_{23}(q_{32}-3q_{31})\\
& & \hskip2cm + (q_{12}q_{21} + q_{13}q_{23}) (q_{31}+q_{32}) \biggr). \nonumber
\eea
For simplicity arguments, one can further impose the conditions: $q_{12} = q_{23} = q_{31}$, $q_{21} = q_{13} = q_{32}$ and $q_{12} \neq \pm q_{21}$ which lead to $d_1 = d_2 = d_3 > 0$ and ${\cal Q}^{-1} = (q_{12} + q_{21})^2 \neq 0$. For the choice $q_{12} = 1$ and $q_{21} = 0$, one simply has ${\cal Q} = 1$. After setting the two heavier moduli at their minimum, i.e. $\langle \phi^2 \rangle = 0 = \langle \phi^3 \rangle$, the single field effective inflationary potential for $\phi^1$ modulus takes the following form,
\bea
\label{eq:V-phi1-global}
& & \hskip-1.5cm V(\phi^1) = -\, {\cal B} \, e^{-3\sqrt{\frac{3}{2}} \phi^1} \left(\sqrt{\frac{3}{2}} \phi^1 - \frac{3}{2}\, e^{\sqrt{\frac{3}{2}} \phi^1}\, a_1\, \, -a_2 + \frac{1}{3} \right) \\
& & + \frac{15}{4}\, n_0^2\,{\cal C}_2 \, {\cal C}_w \,e^{-5\sqrt{\frac{2}{3}} \phi^1} + 3\, n_0^{3/2}\,{\cal C}_3 \, e^{-\frac{11}{\sqrt6} \phi^1}, \quad {\cal B} = -\, 2\, \hat\eta\, n_0^{3/2} \, {\cal C}_1 > 0, \nonumber
\eea
where $a_1 \equiv - \, \frac{(d_1\, d_2\, d_3)^{1/3}}{n_0^{3/2}\, \hat\eta \, {\cal C}_1} \, \geq \,0$, and using (\ref{eq:V-task2-3}), all the moduli VEVs can be determined as,
\bea
\label{eq:d-vs-phi1}
& & \hskip-1cm d_1 = d_2 = d_3 = - \hat\eta \, {\cal C}_1\, n_0^{3/2}\, e^{-\sqrt{\frac{3}{2}} \langle\phi^1\rangle} \left(\sqrt{\frac{3}{2}} \, \langle \phi^1 \rangle -a_2\right) +\frac{25}{12} \, n_0^2 \, {\cal C}_2 \, {\cal C}_w \,e^{-2\sqrt{\frac{2}{3}} \langle\phi^1\rangle} \\
& & \hskip-0.5cm  + \frac{11}{6} \, n_0^{3/2} \, {\cal C}_3 \, e^{-\frac{5}{\sqrt6} \langle\phi^1\rangle}, \qquad \langle \phi^2 \rangle = 0 = \langle \phi^3 \rangle, \nonumber
\eea
Note that the leading piece which corresponds to the first line of Eq.~(\ref{eq:V-phi1-global}) involves two parameters $a_1$ and $a_2$. However, introducing another parameter $x$ via $a_1 \equiv e^{-a_2 - 1- x}$ along with a constant shift in the field $\phi^1$ effectively makes it depend on a single parameter $x$ only. The shifted modulus is:
\bea
\label{eq:shiftedphi}
& & \sqrt{\frac{3}{2}}\,  \phi^1  - a_ 2  - 1 \equiv \sqrt{\frac{3}{2}}\, \phi.
\eea
Subsequently, one can rewrite the single-field potential in Eq.~(\ref{eq:V-phi1-global}) in the following form,
\bea
\label{eq:Vinf-global}
& & \hskip-1.5cm V_{\rm inf}(\phi) = -\, \tilde{\cal B} \, e^{-3\sqrt{\frac{3}{2}} \phi} \left(\sqrt{\frac{3}{2}} \phi - \frac{3}{2}\, e^{-x\, + \sqrt{\frac{3}{2}} \phi} + \frac{4}{3} \right) + \tilde{\cal C}_2 \, e^{-5\sqrt{\frac{2}{3}} \phi} + \,\tilde{\cal C}_3 \, e^{-\frac{11}{\sqrt6} \phi},
\eea
where the various coefficients depending on the model dependent parameters $g_s$, $|W_0|$ and $\lambda$ are given as below,
\bea
\label{eq:tildeBC2C3}
& & \tilde{\cal B} \equiv \tilde{\cal B}(|W_0|,g_s) = -\,\kappa \frac{\chi({\rm CY})\, \sqrt{g_s}\, \, |W_0|^2\,e^{-10-\frac{9 \zeta[3]}{g_s^2\, \pi^2}}}{64\pi} > 0,\\
& & \tilde{\cal C}_2 = \frac{15}{4}\, \kappa \, {\cal C}_w\,|W_0|^2 n_0^{1/3}\, e^{-\frac{100}{9}-\frac{10 \zeta[3]}{g_s^2\, \pi^2}}, \qquad \tilde{\cal C}_3 = -\frac{72\,\kappa^2\, \lambda\, |W_0|^4\, }{g_s^{3/2}\,n_0^{1/3}}\, e^{-\frac{110}{9}-\frac{11 \zeta[3]}{g_s^2\, \pi^2}},\nonumber
\eea
where $\kappa \equiv e^{K_{cs}} g_s/(8 \pi) = 1$ and $\lambda$ is typically given as $|\lambda| \simeq {\cal O}(10^{-4}-10^{-3})$ \cite{Grimm:2017pid,Cicoli:2023njy}. Further, the derivatives and the Hessian take the following respective forms,
\bea
\label{eq:derivatives-Vinf-global}
& & \hskip-1.5cm \partial_{\phi} V_{\rm inf} = \frac{3\sqrt{3}}{\sqrt{2}}\, \tilde{\cal B} \, e^{-3\sqrt{\frac{3}{2}} \phi} \left(\sqrt{\frac{3}{2}} \phi - \, e^{-\,x\,+ \sqrt{\frac{3}{2}} \phi}\, + 1 \right) -\, 5\sqrt{\frac{2}{3}} \tilde{\cal C}_2 \, e^{-5\sqrt{\frac{2}{3}} \phi} -\frac{11}{\sqrt6} \,\tilde{\cal C}_3 \, e^{-\frac{11}{\sqrt6} \phi}, \\
& & \hskip-1.5cm \partial^2_{\phi} V_{\rm inf} = -\,\frac{27}{2}\, \tilde{\cal B} \, e^{-3\sqrt{\frac{3}{2}} \phi} \left(\sqrt{\frac{3}{2}} \phi - \frac{2}{3}\, e^{- x\,+ \sqrt{\frac{3}{2}} \phi}\, + \frac{2}{3} \right) + \, \frac{50}{3} \tilde{\cal C}_2 \, e^{-5\sqrt{\frac{2}{3}} \phi} +\frac{121}{6} \,\tilde{\cal C}_3 \, e^{-\frac{11}{\sqrt6} \phi}. \nonumber
\eea
The inflationary potential (\ref{eq:Vinf-global}) basically involves a total of four parameters which control the dynamics of the inflaton modulus $\phi$ and can be relevant for realising cosmological observables, with/without the sub-leading corrections. These parameters are: $\{x, \, \tilde{\cal B}, \, \tilde{\cal C}_2, \, \tilde{\cal C}_3\}$, where we recall that $\tilde{B}$ controls the leading order BBHL and log-loop effects while $\tilde{\cal C}_2$ parameter controls the winding-loop effects, and the $\tilde{\cal C}_3$ parameter determines the higher derivative F$^4$-corrections. In addition, the parameter $x$ introduced via $a_1 \equiv e^{-a_2 - 1- x}$ controls the uplifting and solely determines the VEV of the $\phi$ modulus in the absence of sub-leading effects, and therefore it also controls the inflaton shift during inflation. These four parameters $\{x, \, \tilde{\cal B}, \, \tilde{\cal C}_2, \, \tilde{\cal C}_3\}$ generically depend on the various model dependent `stringy ingredients' such as $\left\{g_s,\, W_0,\, \chi({\rm CY}), \, n_0, \,{\cal C}_w, \, \lambda\right\}$ as seen from Eq.~(\ref{eq:tildeBC2C3}). For our model we consider following model dependent parameters,
\bea
\label{eq:global-model1}
& & \hskip-0.3cm \chi({\rm CY}) = -224, \quad n_0 =2, \quad g_s = \frac{1}{3}, \quad x = 10^{-4},
\eea
which using (\ref{eq:def-xi-eta}) results in the following,
\bea
\label{eq:global-model2}
& & \hskip-2cm \hat\xi = 2.82024, \quad \hat\eta = -0.428811, \quad a_2 = 5.96834, \quad a_1 \equiv e^{-a_2 -x -1} = 0.00094112. \nonumber 
\eea
Subsequently, using the string parameters as in (\ref{eq:global-model1}) further results in,
\bea
\label{eq:global-model3}
& & \tilde{\cal B} = 1.51694 \times 10^{-9}\, |W_0|^2, \qquad \tilde{\cal C}_2 = 1.22570 \times 10^{-9}\, {\cal C}_w\, |W_0|^2, \\
& & \tilde{\cal C}_3 = - 8.47389 \times 10^{-9}\, \lambda \, |W_0|^4. \nonumber
\eea
Thus we further need to choose just three parameters, namely $W_0$, ${\cal C}_w$ and $\lambda$ for our model building. Also, for the choice $g_s = 1/3$ which we have set, the ratio of the two coefficients corresponding to the sub-leading corrections included through the coefficients ${\cal C}_2$ and ${\cal C}_3$ are estimated as follows,
\bea
\label{eq:ratio-R1-R2}
& & {\cal R}_1 = \frac{\tilde{\cal C}_2}{\tilde{\cal B}} = 0.80801\, {\cal C}_w, \qquad {\cal R}_2 = \frac{\tilde{\cal C}_3}{\tilde{\cal B}} = -5.58619\,|W_0|^2 \lambda.
\eea
This analysis suggests that one needs ${\cal C}_w \ll 1$ for ensuring control over the winding type string-loop correction, while smaller values for ($W_0^2|\lambda|$) are needed for control against the F$^4$-corrections. However, given that winding corrections generically depend on the CS moduli, one may expect to have a tuned value of the ${\cal C}_w$ parameter, and as argued in \cite{Grimm:2017pid,Cicoli:2023njy} one may expect $\lambda \simeq -10^{-4}$ for typical models, making the viability of the inflationary model robust against F$^4$-corrections as well. These arguments are demonstrated in a numerical model given as below \cite{Bera:2024zsk},
\bea
\label{eq:global-M5}
& & W_0 = 0.038, \qquad {\cal C}_w = 5\cdot10^{-5}, \qquad \lambda = -10^{-4},\\
& & \nonumber\\
& & \tilde{\cal B} = 2.19046 \times 10^{-12}, \qquad \tilde{\cal C}_2 = 8.84958 \times 10^{-17}, \qquad \tilde{\cal C}_3 = 1.76692 \times 10^{-18},\nonumber\\
& & \nonumber\\
& & \langle \phi \rangle = -0.00841545, \quad  \langle \tau_\alpha \rangle = 103.409, \quad \langle {\cal V} \rangle = 743.568, \quad \langle V \rangle = 3.64835 \times 10^{-13},\nonumber\\
& & m_\phi^2 = 0.015697 \, m^2_{\phi^\alpha}, \qquad m^2_{\phi^\alpha} = 6.70767 \times 10^{-12} \quad {\rm for} \quad \alpha \in\{ 2, 3\},\nonumber\\
& & \nonumber\\
& & \phi^\ast = 0.000567702, \quad \epsilon_V^\ast = 7.05464 \times 10^{-7}, \qquad \eta_V^\ast = -0.0199979, \qquad N_e \simeq 97,\nonumber\\
& & P_s \simeq 2.1 \times 10^{-9},\qquad n_s \simeq 0.96, \qquad r \simeq 1.1 \times 10^{-5}. \nonumber
\eea
In fact, the model presented in (\ref{eq:global-M5}) gives similar cosmological predictions even for $\lambda = -10^{-3}$ with $W_0 = 0.0334$. However considering ${\cal C}_w \gtrsim 10^{-4}$, the single-field approximation does not be remain intact as the mass hierarchy between the heavier moduli and the overall volume modulus is not significant.


\subsection{Fibre Inflation}
In the previous model we argued that the volume modulus ${\cal V}$ drives an inflection point inflation while the sub-leading corrections are harmless for the inflationary dynamics, in a given region of the parameter space. Now we aim to show that  after stabilizing the overall volume modulus via the perturbative LVS process, and fixing another direction by introducing the gauge fluxes, one is left with a single field potential which can drive the inflation by the sub-leading effects.

For example, in order to obtain a chiral visible sector on the D7-brane stacks wrapping $D_1$, $D_2$ and $D_3$, we need to consistently turn on worldvolume gauge fluxes of the form: ${\cal F}_i = \sum_{j=1}^{h^{1,1}} f_{ij}\hat{D}_j + \frac12 \hat{D}_i - \iota_{D_i}^*B$
with $f_{ij}\in \mathbb{Z}$. Here, the half-integer contribution is due to Freed-Witten anomaly cancellation \cite{Minasian:1997mm,Freed:1999vc}. However, given that the three stacks of D7-branes are wrapping a spin divisor K3 with $c_1({\rm K3}) = 0$, and given that the triple intersections on this CY are even, the pullback of the $B$-field on any divisor $D_\alpha$ does not generate a half-integer flux, and therefore one can appropriately adjust fluxes such that ${\cal F}_\alpha \in {\mathbb Z}$ for all $\alpha \in \{1, 2, 3\}$. We shall therefore consider a non-vanishing gauge flux ${\cal F}_3$ on the worldvolume of the $D_3$ divisor while considering ${\cal F}_1 = 0 = {\cal F}_2$. Subsequently, similar to \cite{Cicoli:2017axo}, the vanishing of FI parameter leads to
\bea
\label{eq:tau2=tau3}
& & \tau_1= q \, \tau_2, \quad {\rm where} \quad q = -q_{31}/q_{32}.
\eea
Using 
$\tau_1 = \tau_2$  for $q=1$ in (\ref{eq:tau2=tau3}) one is left with two K\"ahler moduli, and the scalar potential (\ref{eq:Vfinal-simp}) takes the following form,
\bea
\label{eq:Vfinal-simp1}
& & \hskip-1cm V_{\rm tot} = V_{\rm up} + \, {\cal C}_1 \, \left(\frac{\hat\xi - 4\,\hat\eta + 2\,\hat\eta \, \ln{\cal V}}{{\cal V}^3}\right) +  \frac{{\cal C}_3}{{\cal V}^3}\,\left(\frac{2}{\tau_1}+\frac{1}{\tau_3} \right)\\
& &  \hskip-0.5cm - \frac{{\cal C}_2}{{\cal V}^4} \left[\left(C_{w_1}+C_{w_2}+\frac14C_{w_4}\right)\tau_1 + C_{w_3}\tau_3 +  \frac{(C_{w_5}+C_{w_6})\tau_1 \tau_3}{2(\tau_1 + \tau_3)}\right],  \nonumber
\eea
where $\tau_1 = \frac{\sqrt{n_0} {\cal V}}{\sqrt{\tau_3}}$. Using some appropriate uplifting piece, the overall volume ${\cal V}$ can be fixed via perturbative LVS at leading order, resulting in a de-Sitter minimum. Subsequently, using (\ref{eq:Vfinal-simp1}) and considering $\tau_3 \equiv \tau_f = e^{2\varphi/\sqrt{3}}$ for the canonical field $\varphi$ and making the shift $\varphi = \langle \varphi \rangle + \phi$, one gets the following effective single field scalar potential for the shift modulus $\phi$,
\bea
\label{eq:Vinf}
& & \hskip-0.9cm V = {\cal C}_0 \biggl({\cal C}_{\rm up} + {\cal R}_0 e^{-\gamma\phi} - e^{-\frac{\gamma}{2}\phi} +{\cal R}_1 e^{\frac{\gamma}{2}\phi} + {\cal R}_2 e^{\gamma\phi}\biggr),
\eea
where $\gamma=2/\sqrt{3}$ and the other parameters are:
\bea
\label{eq:Cis-new}
& & \hskip-0.5cm {\cal C}_0 =\frac{\sqrt{2}{\cal C}_2 \tilde{\cal C}_w}{\langle{\cal V}\rangle^3 e^{\frac{\gamma}{2}\langle\varphi \rangle}},  \quad {\cal R}_0 = \frac{{\cal C}_3 e^{-\frac{\gamma}{2}\langle\varphi \rangle}}{\sqrt{2}{\cal C}_2\tilde{\cal C}_w}, \quad \frac{{\cal R}_1}{{\cal R}_0} = \frac{\sqrt{2} e^{\sqrt{3}\langle\varphi\rangle}}{\langle{\cal V} \rangle}, \\
& & \hskip-0.5cm \frac{{\cal R}_2[\phi]}{{\cal R}_0} = \frac{{\cal C}_2 |{\cal C}_{w3}| e^{2\gamma\langle\varphi \rangle}}{{\cal C}_3\,\langle{\cal V} \rangle} \biggl[1+\hat{\cal C}_w \left(1+\frac{e^{\sqrt{3}(\phi+\langle\varphi\rangle)}}{\langle{\cal V}\rangle \sqrt{2}}\right)^{-1}\biggr],\nonumber
\eea
where $\tilde{\cal C}_w =(4{\cal C}_{w1} + 4{\cal C}_{w2} + {\cal C}_{w4})/4$ and $\hat{\cal C}_w = -({\cal C}_{w5}+{\cal C}_{w6})/({2|{\cal C}_{w3}|})$. Further, we note that ${\cal C}_0, {\cal R}_0$ and ${\cal R}_1$ do not depend on the inflaton $\phi$ while ${\cal R}_2[\phi]$ exhibits a sub-leading dependence. Subsequently, ${\cal C}_{\rm up} = 1-{\cal R}_0-{\cal R}_1-{\cal R}_2[0]$ is the uplifting required to obtain a dS minimum with small cosmological constant. Let us emphasize  the following points:
\begin{itemize}
\item 
The first three terms of the inflationary potential in Eq.~(\ref{eq:Vinf}) determine the minimum while the other two terms create the steepening. In fact, the first three terms correspond to Starobinsky type inflationary potential \cite{Starobinsky:1980te} (see  also~\cite{Brinkmann:2023eph}). Therefore we need to examine if this inflation remains robust against the other sub-leading corrections and if there are any knock-on effects on inflation dynamics.

\item 
As seen from (\ref{eq:Cis-new}), the parameters ${\cal R}_1 \ll 1$ and ${\cal R}_2 \ll 1$ given that they are volume (${\cal V}$) suppressed as compared to ${\cal R}_0$, and this is to be exploited in finding a sufficiently long plateau.

\item 
The slow-roll parameters do not depend on ${\cal C}_0$ as it is an overall factor seen from (\ref{eq:Vinf}), hence $n_s$, $r$ and $N_e$ can be determined purely by three parameters ${\cal R}_0, {\cal R}_1$ and ${\cal R}_2$, and then ${\cal C}_0$ can be appropriately chosen to match the scalar perturbation amplitude $P_s$.

\item 
Although ${\cal R}_2$ depends on $\phi$, this dependence is suppressed by an extra volume factor, and can be made insignificant for the choice ${\cal C}_{w5} \simeq - {\cal C}_{w6}$, i.e. $\hat{\cal C}_w \simeq 0$. Along these lines, to begin with, ${\cal R}_2$ can be taken as a constant in order to understand the analytics of the leading order dynamics. Under this assumption, ${\cal R}_0 = (1+{\cal R}_1 + 2{\cal R}_2)/2 \simeq 1/2$ for setting the shift $\langle\phi\rangle = 0$, i.e. when $\varphi$ modulus reaches its minimum. 

\end{itemize}

\noindent
Based on these arguments, one can study the inflationary dynamics of the potential (\ref{eq:Vinf}) and determine the range of ${\cal R}_1$ and ${\cal R}_2$ that could produce a long enough plateau to generate sufficient efoldings $N_e$ and suitable $(n_s, r)$ values, and then ${\cal C}_0$ is fixed by matching the scalar perturbation amplitude $P_s$. A set of benchmark numerical parameters presenting the important features of this model is given as below
\bea
\label{eq:benchmark4}
& & \hskip-0.5cm {\cal C}_0 = 5.5 \cdot 10^{-10}, \quad {\cal R}_1 = 5\cdot10^{-5}, \quad {\cal R}_2 = 10^{-7},\\
& & \hskip-0.5cm \phi_{\rm end} \simeq 1.03, \quad \phi^\ast \simeq 6.37, \quad N_e \simeq 50.6, \nonumber\\
& & \hskip-0.5cm P_s \simeq 2.1 \cdot 10^{-9}, \quad n_s \simeq 0.967, \quad r \simeq 0.0085. \nonumber
\eea

\noindent
{\bf Estimating the stringy parameters:} So far we have studied the inflationary dynamics of the potential (\ref{eq:Vinf}) simply considering the ${\cal C}_0, {\cal R}_0, {\cal R}_1$ and ${\cal R}_2$ as constant parameters, without going into the details of the corresponding stringy parameters. Now we show that, using the various stringy parameters, similar numerical models can be produced. For that purpose, using (\ref{eq:Cis-new}) the values of stringy parameters can be typically estimated by  the following relations:
\bea
\label{eq:stringy-parameters-gen}
& & \hskip-0.5cm |\hat{W}_0| \equiv e^{\frac{1}{2}K_{\rm cs}}\,|W_0| = \frac{2^{2/3}\, {\cal C}_0^{1/4}\, \sqrt{\pi}\, {\cal R}_0^{1/12}\, {\cal R}_1^{1/6}\, {\langle{\cal V} \rangle}^{11/12}}{3^{1/4}\, g_s^{1/8}\, |\lambda|^{1/4}},\\
& & \hskip-0.5cm \tilde{\cal C}_w = \frac{2\sqrt{3} {\cal C}_0^{1/2} {\langle{\cal V} \rangle}^{3/2} \sqrt{|\lambda|}}{g_s^{3/4}\, \sqrt{{\cal R}_0}}, \quad {\cal C}_{w3} = \frac{2 {\cal R}_0{\cal R}_2}{{\cal R}_1} \tilde{\cal C}_{w}, \quad \langle{\cal V} \rangle = \frac{\sqrt{2}\, {\cal R}_0\, e^{\sqrt{3}\langle\varphi\rangle}}{{\cal R}_1}, \nonumber
\eea
where we recall that ${\cal R}_0 = (1+{\cal R}_1 + 2{\cal R}_2)/2 \sim 1/2$ under the assumption that ${\cal R}_0, {\cal R}_1$ and ${\cal R}_2$ are constant parameters where, in addition, being volume suppressed one anticipates that ${\cal R}_1 \ll 1$ and ${\cal R}_2 \ll 1$ are natural choices. However, the numerical estimates from (\ref{eq:benchmark4})-(\ref{eq:stringy-parameters-gen}) also show that the volume $\langle{\cal V}\rangle \simeq 10^3-10^4$  corresponds to negative values of $\langle \varphi\rangle$, i.e. fractional VEVs for the $\tau_f$ modulus, which can still be sufficient to trust the overall effective supergravity analysis if $\tau_{\rm str}^{1/4} = (\tau_E \, g_s)^{1/4} \gg \frac{1}{2\pi}$, where the string-frame divisor volume and the Einstein-frame divisor volume are related via $\tau_E = \tau_{\rm str}/g_s$ \cite{Cicoli:2017shd,Cicoli:2017axo,AbdusSalam:2020ywo}. Now, we present the following numerical model:
\bea
\label{eq:final-model}
& & \hskip-0.5cm |\hat{W_0}| = 6, \quad {\tilde{\cal C}_w} = 7, \quad |{\cal C}_{w3}| = 1/2, \quad \hat{\cal C}_w = 1, \quad |\lambda| = 1.27\cdot10^{-2}\\
& & \hskip-0.5cm g_s = 0.25, \quad \langle\varphi\rangle = -3, \quad \tau_{\rm str}^{1/4} \simeq 0.3 > \frac{1}{2\pi},   \quad \langle{\cal V}\rangle \simeq 3567,\nonumber
\eea
which result in ${\cal C}_0 \simeq 4.42 \cdot 10^{-10}, \, \, {\cal R}_0 \simeq 0.5, \, \, {\cal R}_1 = 1.1\cdot10^{-6}$, and 
\bea
{\cal R}_2[\phi] = \frac{0.142857 + 7.84174 \cdot 10^{-8} e^{\sqrt{3} \phi}}{910877 + e^{\sqrt{3} \phi}}.
\eea
Subsequently the cosmological predictions are:
\bea
& & \phi_{\rm end} \simeq 1.03, \quad \phi^\ast \simeq 6.47, \quad N_e \simeq 55.1, \\
& & P_s^\ast \simeq 2.13\cdot10^{-9}, \quad n_s^\ast \simeq 0.967, \quad r^\ast \simeq 6.7\times10^{-3}.\nonumber
\eea
For the model given in (\ref{eq:final-model}), the effective supergravity approximations are justified by ensuring the mass hierarchies given as below,
\bea
\label{eq:mass-hierarchy-num}
& & \hskip-0.5cm M_s = \frac{\sqrt{\pi} \, g_s^{1/4}}{{\sqrt{\cal V}}} \sim 2.1\cdot 10^{-2}, \quad M_{\rm KK}\simeq \frac{\sqrt{\pi}}{\sqrt{\cal V} \, \tau_{\rm bulk}^{1/4}} \sim 3.8 \cdot10^{-3}, \\
& & \hskip-0.5cm m_{3/2} = e^{\frac{1}{2}K}\, |W_0| \simeq \sqrt{\frac{g_s}{8 \pi}} \frac{|\hat{W_0}|}{{\cal V}} \sim 1.7\cdot10^{-4}, \nonumber\\
& & \hskip-0.5cm m_{\rm inf} \sim 2.9\cdot10^{-6}, \quad H^\ast \simeq \sqrt{\frac{V_{inf}^\ast}{3}} \sim 8.4\cdot10^{-6}, \nonumber
\eea
where for $M_{\rm KK}$ for the bulk modulus $\tau_{\rm bulk} \sim {\cal V}^{2/3}$ \cite{Cicoli:2017axo}, and all the masses in (\ref{eq:mass-hierarchy-num}) are expressed in units of $M_p$. Here we note that $m_{3/2} < M_{\rm KK}$ results in a bound on $|\hat{W}_0|$ which is given as: $\sqrt{\frac{\kappa}{\pi}} \, |\hat{W}_0| < {\cal V}^{1/3}$ where $\kappa = g_s/(8\pi)$ \cite{Cicoli:2013swa,AbdusSalam:2020ywo}. Such a bound may be hard to satisfy for models with large $|\hat{W}_0|$ values and having the smaller volumes!


\section{Summary and Conclusions}
\label{sec_summary}

In conclusion the Large Volume Scenario (LVS) manifests in two distinct formulations:
Firstly, via the standard LVS, which combines BBHL corrections to the K\"ahler potential with non-perturbative superpotential corrections and, secondly, through the perturbative LVS, which incorporates string-loop effects (of log-loop type) alongside BBHL corrections.
Both approaches stabilize the overall volume modulus ${\cal V}$  at exponentially large values, corresponding to a non-supersymmetric AdS minimum. This minimum can subsequently be uplifted to a de Sitter vacuum through appropriate mechanisms.
Building upon the global embedding program for perturbative LVS initiated in \cite{Leontaris:2022rzj}, this report reviews two inflationary scenarios realized within this framework \cite{Bera:2024ihl, Bera:2024zsk}. 


The first one is the ``volume-modulus inflation" or the ``inflection point inflation" which is a small field model of inflation driven by an effective potential induced by the BBHL and log-loop corrections to the K\"ahler potential. We discussed the global embedding of the inflationary scenario proposed in \cite{Antoniadis:2020stf} by using an explicit K3-fibred CY orientifold with properties similar to those of the toroidal model. Subsequently, we studied the robustness of the stabilization procedure and that of the inflationary dynamics against various possible corrections which are inevitable once the global model is chosen/fixed. These corrections are winding-type string loop corrections and the higher derivative F$^4$ correction \cite{Bera:2024zsk}.

The second model we discussed is popularly known as ``fibre inflation" \cite{Cicoli:2008gp} which is a large field model of inflation developed in the framework of standard LVS. However, fibre inflation in the standard LVS framework typically faces a field-range bound imposed by K\"ahler cone conditions~\cite{Cicoli:2018tcq}. This constraint originates from the rigid exceptional divisor—an essential component of standard LVS that simultaneously enforces two key features: (i) the Swiss-cheese structure of the Calabi-Yau volume form, and (ii) non-perturbative superpotential contributions.
Since perturbative LVS does not require such exceptional divisors, it naturally circumvents this field-range problem, offering a theoretically consistent framework for large-field inflation \cite{Bera:2024ihl}.

As we conclude, let us note that all the string-loop corrections used for these models, i.e. the KK/Winding type \cite{Berg:2004sj,vonGersdorff:2005bf,Berg:2005ja,Berg:2007wt,Cicoli:2007xp} and the lop-loop type \cite{Antoniadis:2018hqy,Antoniadis:2018ngr,Antoniadis:2019doc,Antoniadis:2019rkh,Antoniadis:2020ryh,Antoniadis:2020stf}, are motivated by the toroidal results. Although some of these corrections have been recently (re-)derived via a field theoretic approach \cite{Gao:2022uop}, a direct computation of these corrections for the CY orientifold case is still missing, and  further work is required to guarantee the viability of such inflationary models. 


\subsection*{Acknowledgments}
\noindent
PS is thankful to the {\it Department of Science and Technology (DST), India} for the kind support.



\bibliographystyle{utphys}
\bibliography{reference}

\end{document}